\documentstyle[12pt]{article}
\begin{document}
\begin{flushright}
hep-ph/9801210\\[10mm]
\end{flushright}
\begin{center}
{\Large {\bf $\rho^0-\omega$ mixing in ${\rm U(3)}_L\times {\rm U(3)}_R$
chiral theory of mesons}}\\[5mm]
Dao-Neng Gao\\
{\small Center for Fundamental Physics, University of Science and Technology of China}\\
{\small Hefei, Anhui 230026, People's Republic of China}\\[2mm]
Mu-Lin Yan\\
{\small Chinese Center for Advanced Science and Technology (World Lab)\\
P.O.Box 8730, Beijing, 100080, People's Republic of China}\\
{\small and}\\
{\small Center for Fundamental Physics,
University of Science and Technology of China\\
Hefei, Anhui 230026, People's Republic of China}\footnote{mailing address}
\end{center}
\vspace{8mm}
\begin{abstract}
\noindent
In this paper, U(3)$_L\times$U(3)$_R$ chiral theory of mesons is extended
to the leading order in quark mass expansion in order to evaluate the
$\rho^0-\omega$ mixing. It is shown that the use of path integral method
to integrate out the quark fields naturally leads to the $\rho^0-\omega$
mixing vertices, and this mixing is entirely from the quark loop
in this theory. The off-shell behaviour of the mixing amplitude is analyzed.
The on-shell mixing amplitude is obtained from the decay
$\omega\rightarrow \pi^+\pi^-$. Furthermore,
the constraints on the light quark mass parameters are extracted from
the $\rho^0-\omega$ mixing and the mesons masses, and the mass splitting
of $K^*(892)$-mesons is predicted.
\end{abstract}

\vspace{4mm}

\newpage
\section{Introduction}

In the limit of vanishing light quark masses, the lagrangian of quantum
chromodynamics (QCD) possesses the exact chiral SU(3)$_L\times$ SU(3)$_R$
symmetry. It has been known that this symmetry is spontaneously
broken to SU(3)$_V$ with the appearances of eight Goldstone pseudoscalar
particles ($\pi, K, \eta$) which dominate low energy dynamics of
the strong interaction.
Chiral perturbation theory (ChPT), which is expanded in powers of
derivatives of the mesons fields, is rigorous and phenomenologically
successful in describing the physics of the pseudoscalar mesons
at very low energies \cite{SGL}. On the other hand, chiral symmetry is
explicitly broken due to the small current quark masses, which
leads to the nonzero masses of the pseudoscalar mesons.
In addition, the inequality of the light quark masses, especially, $m_u\ne m_d$,
does break the isospin symmetry or charge symmetry. This breaking of isospin
symmetry induces various hadron mixings such as $\pi^0-\eta$, $\rho^0-\omega$,
and $\Lambda-\Sigma^0$ mixings etc \cite{MNS}. In this paper, we will focus on the $\rho^0-\omega$
mixing, which is considered as the important source of charge symmetry breaking
in nuclear physics.

This investigation of $\rho^0-\omega$ mixing has been an active subject
 [2--14],
and the mixing amplitude for on-mass-shell vector mesons has been
observed directly in the measurement of the pion form-factor in the time-like
region from the process $e^+ e^-\rightarrow \pi^+\pi^-$ \cite{Bark}.
For roughly twenty years, $\rho^0-\omega$ mixing amplitude was assumed
constant or momentum independent, even if $\rho$ and $\omega$ mesons have
the space-like momenta, far from the on-shell point.
Several years ago, this assumption was firstly questioned by
Goldman, Henderson, and Thomas, and the mixing amplitude was found
to be significantly momentum dependent within a simple
quark loop model \cite{GHT}.
Subsequently, various authors have argued such $q^2$ dependence
of the $\rho^0-\omega$ mixing amplitude (where $q^2$ denotes the four-momentum square
of the vector mesons)
by using various theoretical approaches \cite{Momenta, OC1}.
In particular, the authors of Ref. \cite{OC1} has pointed out that
$\rho^0-\omega$ mixing amplitude must vanish at $q^2=0$ within a broad class
of models.

The purpose of the present paper is to study the $\rho^0-\omega$
mixing amplitude in the framework of U(3)$_L\times$U(3)$_R$
chiral theory of mesons \cite{Li1,Li2}.
This theory could be regarded as a realization of
current algebra, chiral symmetry, and vector meson dominance (VMD).
The meson fields including pseudoscalar, vector, and axial-vector mesons
are introduced into the present theory as the bound state of quark fields.
The effective lagrangian is obtained by using the path integral method to
integrate out the quark fields, and the kinetic terms of the mesons are
generated from quark loop naturally (see Refs. \cite{Li1,Li2} for details).
The present theory has been investigated extensively, and the theoretical
results agree
with the experimental data well
\cite{Li3,Li4,GLY,WY,Li5}. Particularly, in Ref. \cite{WY},
starting from the U(3)$_L\times$U(3)$_R$ chiral fields
theory of mesons, and by using path integration method to integrate out
the vector and axial-vector resonances,
the authors have derived the chiral coupling
constants of ChPT ($L_1, L_2, L_3, L_9$ and $L_{10}$).
The results are in good agreement with
the experimental values of the $L_i$ at $\mu=m_\rho$ in ChPT. Therefore,
the QCD constraints discussed
in Ref. \cite{EGLPR} are met by this theory.

It has been known that $\rho^0-\omega$ mixing amplitudes receive the contributions from
two sources: isospin symmetry breaking due to $u$-$d$ quark mass difference and
electromagnetic interactions. As mentioned in Ref. \cite{Li1},
VMD \cite{KLZ,Sak} in the meson physics is natural consequence
of the present theory instead of an input. Therefore, the dynamics of the electromagnetic
interactions of mesons has been well introduced and established. Thus,
the calculation of $\rho^0-\omega$ mixing amplitude from the
transition $\rho\rightarrow \gamma \rightarrow \omega$ is straightforward.
In Refs. \cite{Li1,Li2}, the light quark masses are set to be massless.
When the current quark mass terms, which explicitly break
the chiral symmetry, are included in the present theory,
the use of path integral method to integrate out the quark fields will
naturally reduce the $\rho^0-\omega$ mixing terms in the mesons effective
lagrangian (see below).
Therefore, at the leading order in quark mass expansion and $O(\alpha_{\rm EM})$, the amplitude of
$\rho^0-\omega$ mixing can be evaluated systematically
within U(3)$_L\times$U(3)$_R$ chiral theory of mesons.  As will be shown below,
the $\rho^0-\omega$ mixing receives the contributions entirely
from the quark loop in the framework of the present theory.

In Ref. \cite{UR}, in order to calculate the $\rho^0-\omega$ mixing,
the author extended the chiral couplings of the low-lying vector
resonances in chiral
perturbation theory \cite{Ecker} to a lagrangian that contains two
vector fields, and made some assumptions to pin down the coupling
constants of the lagrangian related to this mixing.

The contents of the paper are organized as follows.
In Sec. 2, we present the basic notations of the U(3)$_L\times$U(3)$_R$ chiral theory of
mesons, and extend the theory to the leading order in quark mass expansion
in order to derive the $\rho^0-\omega$ mixing due to isospin symmetry breaking.
In Sec. 3, the off-shell and on-shell $\rho^0-\omega$ mixing amplitude are
studied, and the constraints on the light quark mass parameters are extracted.
In Sec. 4,  we give a summary of the results.

\section{U(3)$_L\times$U(3)$_R$ chiral theory of mesons}

The basic lagrangian of U(3)$_L\times$U(3)$_R$ chiral theory of mesons
is (hereafter we use the notation of Refs. \cite{Li1,Li2})
\begin{eqnarray}
\lefteqn{{\cal L}=\bar{\psi}(x)(i\gamma\cdot\partial+\gamma\cdot v
+e_0Q\gamma\cdot A+\gamma\cdot a\gamma_{5}-M
-mu(x))\psi(x)}\nonumber \\
& &+{1\over 2}m^{2}_{1}(\rho^{\mu}_{i}\rho_{\mu i}+
\omega^{\mu}\omega_{\mu}+a^{\mu}_{i}a_{\mu i}+f^{\mu}f_{\mu})\nonumber\\
& &+{1\over 2}m^2_2(K_{\mu}^{*a}K^{*a\mu}+K_1^{\mu}K_{1\mu})\nonumber \\
& &+{1\over 4}m^2_3(\phi_{\mu} \phi^{\mu}+f_s^{\mu}f_{s\mu})
+{\cal L}_{\rm EM}
\end{eqnarray}
with
\begin{eqnarray}
& &u(x)={\rm exp}[i\gamma_{5} (\tau_{i}\pi_{i}+\lambda_a K^a+\eta
+\eta^{\prime})],\nonumber\\
& &a_{\mu}=\tau_{i}a^{i}_{\mu}+\lambda_a K^a_{1\mu}+(\frac{2}{3}
+\frac{1}{\sqrt{3}}
\lambda_8)f_{\mu}+(\frac{1}{3}-\frac{1}{\sqrt{3}} \lambda_8)f_{s\mu},\nonumber\\
& &v_{\mu}=\tau_{i}\rho^{i}_{\mu}+\lambda_a K_{\mu}^{*a}+(\frac{2}{3}+
\frac{1}{\sqrt{3}}\lambda_8)\omega_{\mu}+(\frac{1}{3}-
\frac{1}{\sqrt{3}}\lambda_8)\phi_{\mu},
\end{eqnarray}
where $i$=1, 2, 3 and $a$=4, 5, 6, 7.
The $\psi$ in Eq.(1) is $u$, $d$, $s$ quark fields.
$M$=diag($m_u$, $m_d$, $m_s$) is the quark mass matrix, which represents the
explicit chiral symmetry breaking in this theory.
$m$ is a parameter related to the quark condensate.
${\cal L}_{\rm EM}$ is the lagrangian of the electromagnetic interaction.
$A_\mu$ is the photon field, and $Q$ is the electric charge operator of the
quark fileds.
Note that there are no kinetic terms in eq. (1) for the meson fields
including pseudoscalar ($\pi, K, \eta$), vector ($v_\mu$),
and axial-vector ($a_\mu$) because they are composed fields of quark fields
instead of the fundamental fields.
The kinetic terms for these fields will be generated from quark loops.

Following Ref. \cite{Li1}, the effective lagrangian of mesons (indicated
by ${\cal M}$) are obtained by performing path integrations over the quark
fields,
\begin{equation}
{\rm exp}\{i\int d^4 x {\cal L^M}\}=
\int [d\psi][d\bar{\psi}]{\rm exp}\{i\int d^4 x {\cal L}\}.
\end{equation}
Using the dimensional regularization, and in the chiral limit, i.e. light quark
masses are zero, the effective lagrangian ${\cal L}_{RE}$ (normal parity
part) and ${\cal L}_{IM}$ (abnormal parity part) has been evaluated
in Refs. \cite{Li1, Li2}. Here we give the  first two terms
of ${\cal L}_{RE}$ in order to present the notations explicitly.
\begin{eqnarray}
{\cal L}_{RE}&=&\frac{N_C}{(4\pi)^2}m^2\frac{D}{4}\Gamma(2-\frac{D}{2})
Tr\hspace{0.01in}D_{\mu}UD^{\mu}U^\dagger\nonumber \\
& &-\frac{1}{3}\frac{N_c}{(4\pi)^2}
\frac{D}{4}\Gamma(2-\frac{D}{2})Tr(v_{\mu\nu}v^{\mu\nu}+a_{\mu\nu}a^{\mu\nu})
+...,
\end{eqnarray}
where $U={\rm exp}i(\tau_i\pi_i+\lambda_a K^a+\eta+\eta^\prime)$, and
\begin{eqnarray*}
D_{\mu} U=\partial_\mu U-i[v_{\mu}, U]+i\{a_\mu, U\},\\
D_{\mu} U^\dagger=\partial_\mu U^\dagger-i[v_{\mu}, U^\dagger]-i\{a_\mu, U^\dagger\},\\
v_{\mu\nu}=\partial_\mu v_\nu-\partial_\nu v_\mu-i[v_\mu,v_\nu]-i[a_\mu, a_\nu],\\
a_{\mu\nu}=\partial_\mu a_\nu-\partial_\nu a_\mu-i[a_\mu,v_\nu]-i[v_\mu, a_\nu].
\end{eqnarray*}
It is obvious that quark loop integral gives rise to the divergences in the
effective lagrangian [eq. (4)].
In Refs. \cite{Li1,Li2}, in order to build a physical effective meson theory,
a universal coupling constant $g$ has been introduced
\begin{eqnarray}
\frac{F^2}{16}=\frac{N_c}{(4\pi)^2}m^2\frac{D}{2}\Gamma(2-\frac{D}{2}),\\
g^2=\frac{8}{3}\frac{N_c}{(4\pi)^2}\frac{D}{4}\Gamma(2-\frac{D}{2})=
\frac{1}{6}\frac{F^2}{m^2}.
\end{eqnarray}
and
\begin{eqnarray}
\frac{F^2}{f_\pi^2}(1-\frac{2c}{g})=1,\\
c=\frac{f_\pi^2}{2gm_\rho^2},
\end{eqnarray}
where $f_\pi$ is the decay constant of pions.

It is straightforward to extend the effective lagrangian
of the present theory to the
leading order in quark mass expansion. Starting from eq. (1), in which the quark mass terms are
included, and using the similar procedure presented in Ref. \cite{Li1},
we can get the effective lagrangian beyond the chiral limit.
The nonzero quark masses will yield
other terms in addition to the effective lagrangian in the chiral limit. At the
leading order in quark mass expansion, the masses of the pseudoscalar mesons
will be no longer zero,
\begin{eqnarray}
m^{2}_{\pi^{+}}=m^2_{\pi^{0}}=-{2\over f^{2}_{\pi}}(m_{u}+m_{d})
\langle 0|\bar{\psi}\psi|0\rangle, \\
m^{2}_{K^{+}}=-{2\over f^{2}_{\pi}}(m_{u}+m_{s})\langle 0|\bar{\psi}\psi|0\rangle,
\\
m^{2}_{K^{0}}=-{2\over f^{2}_{\pi}}(m_{d}+m_{s})\langle 0|\bar{\psi}\psi|0\rangle,
 \\
m^{2}_{\eta}=-{2\over 3f^{2}_{\pi}}(m_{u}+m_{d}+4m_{s})
\langle 0|\bar{\psi}\psi|0\rangle,
\end{eqnarray}
where $\langle 0|\bar{\psi}\psi|0\rangle$ is the quark condensate of the light
flavors \cite{Li1, GLY}.

For the vector mesons, at the leading order in quark mass expansion,
there are no the explicit contributions of their masses. Consequently,
there is no such $\rho^0-\omega$ mixing which is independent of
the momentum (see below).
However, an additional kinetic term of the vector mesons will appear in the
effective lagrangian.
Therefore, to the order of $m_q$, all the kinetic terms of the vector mesons
induced from eqs. (1) and (3) are as follows,
\begin{equation}
{\cal L}^V_{kin}=-\frac{1}{8}g^2 Tr v_{\mu\nu} v^{\mu\nu}+
\frac{2N_c}{3(4\pi)^2 m}Tr\hspace{0.01in}M\hspace{0.01in} v_{\mu \nu}v^{\mu\nu}+
\mbox{higher}\hspace{0.1in} \mbox{order}\hspace{0.1in} \mbox{terms}.
\end{equation}
The second term in eq. (13) contains the $\rho^0-\omega$ mixing due to
isospin symmetry breaking,
which can be easily derived
\begin{equation}
{\cal L}_{\rho\omega}=\frac{1}{4\pi^2}\frac{m_u-m_d}{m}\rho^0_{\mu\nu}
\omega^{\mu\nu}.
\end{equation}
However, the vector meson fields in eq. (13) is not physical fields,
therefore we should make the kinetic terms of the vector mesons fields
in the standard form by redefining these fields.
For $\rho$-mesons, from eq. (13), we have
\begin{equation}
{\cal L}_{kin}^\rho=-\frac{1}{4}g^2(1-\frac{1}{2\pi^2 g^2}\frac{m_u+m_d}{m})
\rho^{i}_{\mu\nu}\rho^{i\mu\nu}.
\end{equation}
The physical $\rho$-mesons field should be defined as
\begin{eqnarray*}
\rho_\mu\longrightarrow \frac{1}{g\sqrt{1-
\frac{1}{2\pi^2 g^2}\frac{m_u+m_d}{m}}}\rho_\mu.
\end{eqnarray*}
Then, the physical mass of $\rho$ mesons is
\begin{equation}
m_\rho^2=\frac{m^2_1}{g^2(1-\frac{1}{2\pi^2 g^2}\frac{m_u+m_d}{m})}=
\frac{m_V^2}{1-\frac{1}{2\pi^2 g^2}\frac{m_u+m_d}{m}}.
\end{equation}
Expanding eq. (16)  to the first order of $m_q$, we obtain
\begin{equation}
m_\rho=m_V(1+\frac{1}{4\pi^2 g^2}\frac{m_u+m_d}{m}),
\end{equation}
here $m_V=\frac{m_1}{g}$ is the vector meson masses in the chiral limit.
Likewise, for $\omega$, $K^*$, and $\phi$ mesons, we can get
\begin{eqnarray}
&&m_\omega=m_V(1+\frac{1}{4\pi^2 g^2}\frac{m_u+m_d}{m}),\\
&&m_{K^{*\pm}}=m_V(1+\frac{1}{4\pi^2 g^2}\frac{m_u+m_s}{m}),\\
&&m_{K^{*0}}=m_V(1+\frac{1}{4\pi^2 g^2}\frac{m_d+m_s}{m}),\\
&&m_{\phi}=m_V(1+\frac{1}{2\pi^2 g^2}\frac{m_s}{m}).
\end{eqnarray}
In the derivation of the above equations, we have set $m_1=m_2=m_3$ in eq. (1)
in order to get
the same vector mesons masses in the chiral limit.
The mass splittings of the vector mesons from the quark mass effect
are reduced [eqs. (17)-(21)] although no explicit vector meson mass terms generated from the
leading order quark mass expansion appear in the effective lagrangian.
Due to the definition of the physical vector mesons,
the $\rho^0-\omega$ mixing of eq. (14) should be translated into
\begin{equation}
{\cal L}_{\rho^0\omega}=\frac{1}{4\pi^2 g^2}\frac{m_u-m_d}{m}\rho^0_{\mu\nu}
\omega^{\mu\nu}.
\end{equation}
The higher order terms in quark mass expansion have been ignored in the
above equation.

\section{$\rho^0-\omega$ mixing amplitude and quark mass parameters}

The $\rho^0-\omega$ mixing induced by the isospin symmetry breaking has
been derived in Sec. 2 [eq. (22)].  Now, we try to get the contribution of
$\rho^0-\omega$ mixing from electromagnetic interactions.
The photon field has been introduced in eq. (1). Following
Refs. \cite{Li1,Li2}, the direct couplings of neutral vector meson
fields ($\rho^0, \omega$, and $\phi$) and the photon fields read
\begin{eqnarray}
{\cal L}_{\rho \gamma}=-\frac{1}{2}{e \over f_\rho}
    \rho_{\mu\nu}^0 (\partial^\mu A^\nu -\partial^\nu A^\mu), \\
{\cal L}_{\omega \gamma}=-\frac{1}{2}{e \over f_\omega}
    \omega_{\mu\nu} (\partial^\mu A^\nu -\partial^\nu A^\mu), \\
{\cal L}_{\phi \gamma}=-\frac{1}{2}{e\over f_\phi}
    \phi_{\mu\nu} (\partial^\mu A^\nu-\partial^\nu A^\mu),
\end{eqnarray}
where
\begin{eqnarray}
\frac{1}{f_\rho}=\frac{1}{2}g,\;\;\;\frac{1}{f_\omega}=\frac{1}{6}g,
\;\;\;\frac{1}{f_\phi}=-\frac{1}{3\sqrt{2}}g.
\end{eqnarray}
Eqs. (23) and (24) will lead to $\rho^0-\omega$ mixing at the order
of $\alpha_{\rm EM}$ through the transition process
$\rho\rightarrow\gamma\rightarrow \omega$, which is
\begin{equation}
{\cal L}_{\rho\omega}=\frac{1}{24}e^2 g^2\rho^0_{\mu\nu}\omega^{\mu\nu}.
\end{equation}
From eqs. (22) and (27), the total $\rho^0-\omega$ mixing is
\begin{equation}
{\cal L}_{\rho\omega}=\frac{1}{4\pi^2 g^2}\frac{m_u-m_d}{m}\rho^0_{\mu\nu}
\omega^{\mu\nu}+\frac{1}{24}e^2 g^2\rho^0_{\mu\nu}\omega^{\mu\nu}.
\end{equation}
Note that the vector mesons mass terms do not lead to
any $\rho^0-\omega$ mixing, therefore, the mixing ${\cal L}_{\rho\omega}$
in the present theory comes entirely from the quark loop. In the standard way,
the two-point Green function associated with $\rho^0-\omega$ mixing is
\begin{eqnarray}
&&\Pi^{\rho\omega}_{\mu\nu}(q^2)=i\int d^4 x\hspace{0.01in}{\rm e}^{iqx}\langle 0|T\hspace{0.01in}\rho_\mu (x)
\omega_\nu(0)\hspace{0.01in} {\rm exp}\{i\int d^4 y\hspace{0.01in}{\cal L}_{\rho\omega}(y)\}|0 \rangle,\nonumber\\
&&=(g_{\mu\nu}-\frac{q_{\mu}q_{\nu}}{q^2})\frac{\theta^{\rho\omega}(q^2)}
{(q^2-m_\rho^2)(q^2-m_\omega^2)},
\end{eqnarray}
where
\begin{equation}
\theta^{\rho\omega}(q^2)=2q^2(\frac{m_u-m_d}{4\pi^2 g^2 m}+
\frac{1}{24}e^2 g^2).
\end{equation}
$\Pi^{\rho\omega}_{\mu\nu}(q^2)$ is of transversal structure,
which is due to the the conservation of the vector currents.
From eq. (30), the off-shell $\rho^0-\omega$ mixing amplitude
$\theta^{\rho\omega}(q^2)$ is obviously momentum dependent, and
vanishes at $q^2=0$. This is consistent with the argument
by O'Connell {\it et al.} in Ref. \cite{OC1} that this mixing amplitude
must vanish at the transition from time-like to space-like four-momentum
within a broad class of models.

Now, we consider the on-shell $\rho^0-\omega$ mixing amplitude. From eq. (30),
we have
\begin{equation}
\theta^{\rho\omega}(m_\rho^2)=2 m_\rho^2(\frac{m_u-m_d}{4\pi^2 g^2 m}+
\frac{1}{24}e^2 g^2).
\end{equation}
The mass difference of $\rho$ and $\omega$ mesons has been ignored here.
Note that the universal coupling constant $g$ has been fixed in this theory
\cite{Li1,Li2,Li3,GLY}, and $m$ can be determined from eqs. (6), (7), and (8).
Therefore, the electromagnetic contribution of the on-shell mixing
amplitude [the second term in eq. (31)] is calculable. By taking $g$=0.39,
$f_\pi$=186 MeV, and $m_\rho$=768.5 MeV,  $\theta^{\rho\omega}(m_\rho^2)$
from  the transition $\rho\rightarrow \gamma\rightarrow \omega$ is
about 0.686$\times 10^{-3}$ GeV$^2$. The $\rho^0-\omega$ mixing leads to
the G-parity forbidden decay of the $\omega$ meson,
$\omega\rightarrow\rho^0\rightarrow \pi^+\pi^-$. By using
the experimental value of the decay width $\Gamma(\omega\rightarrow
\pi^+\pi^-)$, one can get the total on-shell
$\rho^0-\omega$ mixing amplitude \cite{CB}. Therefore, by subtracting the electromagnetic
contribution from the value extracted from this decay, the full
contribution of the mixing amplitude due to isospin symmetry breaking will
be obtained. This could provide an important constraint on the $u$-$d$ quark
mass difference.

Following Refs. \cite{GL2, CB, UR},  the decay width of the process
$\omega\rightarrow\rho^0\rightarrow \pi^+\pi^-$ is expressed as
\begin{equation}
\Gamma(\omega\rightarrow \pi^+\pi^-)=
|\frac{\theta^{\rho\omega}(m_\rho^2)}{m_\omega^2-m_\rho^2-
i(m_\omega\Gamma_\omega-m_\rho\Gamma_\rho)}|^2 \Gamma(\rho\rightarrow \pi^+\pi^-),
\end{equation}
here $\Gamma_\rho$ and $\Gamma_\omega$ are the widths of $\rho$ and $\omega$
mesons respectively, and the decay width of $\rho\rightarrow \pi^+\pi^-$ has been
calculated in Ref. \cite{Li1}
\begin{eqnarray}
&&\Gamma(\rho\rightarrow \pi^+\pi^-)=\frac{f_{\rho\pi\pi}^2}{48\pi}
m_\rho(1-\frac{4m_\pi^2}{m_\rho^2})^{\frac{3}{2}},\nonumber \\
&&f_{\rho\pi\pi}=\frac{2}{g}\{1+\frac{m_\rho^2}{2\pi^2 f_\pi^2}[
(1-\frac{2c}{g})^2-4\pi^2 c^2]\}.
\end{eqnarray}
Using the experimental data $B(\omega\rightarrow \pi^+\pi^-)=2.21\pm0.30\%$ together
with eqs. (32) and (33), we get
\begin{equation}
\theta^{\rho\omega}(m_\rho^2)=-(4.21\pm 0.28)\times 10^{-3}\hspace{0.05in} {\rm GeV}^2,
\end{equation}
which agrees well with the value $-(4.52\pm0.60)
\times 10^{-3}$ GeV$^2$ obtained by Coon and Barrett \cite{CB}.
The error bar in eq. (34) is from the uncertainty in the
branch ratio of the process $\omega\rightarrow \pi^+\pi^-$, and we have neglected
the mass difference between $m_\rho$ and $m_\omega$ and the width
$\Gamma_\omega$ in the denominator of eq. (32). The sign of the on-shell
mixing amplitude has been discussed in Ref. \cite{CSM}, and it is determined from the
relative phase of the $\omega$ to the $\rho$ amplitude in the reaction
$e^+e^-\rightarrow \pi^+\pi^-$ near $m_\rho$ and $m_\omega$.

More recently, the $\rho^0-\omega$ mixing is investigated based on the analysis
of $e^+e^-\rightarrow \pi^+\pi^-$ in Refs. \cite{OC4,OC5}.
Particularly, in Ref. \cite{OC5},
the magnitude, phase and the $s$ dependence of the mixing have been determined
from the pion form-factor in the timelike region,
and the smaller absolute value of the on-shell mixing
amplitude (compared with the value by Coon and Barrett \cite{CB}) has been
obtained. As pointed out by O'Connell {\it et al.} in Ref. \cite{OC3},
the value of the mixing amplitude given in Ref. \cite{CB}
is not the value which provides the optimal fit to the pion
electromagnetic form-factor.

Combining eq. (34)
and eq. (31), we can obtain the $u$-$d$ quark mass difference
\begin{equation}
m_d-m_u=6.14\pm 0.36\hspace{0.05in} {\rm MeV}.
\end{equation}

The value of $m_u-m_d$ can also be extracted from the mass of the mesons.
It has been known that, at the leading order in quark mass expansion,
the mass difference of the non-strange mesons ($\pi$, $a_1$, and $\rho$)
is almost entirely electromagnetic in origin, however, the mass difference
of the strange mesons ($K$, $K_1$, and $K^*$) are from both electromagnetic
interactions and isospin symmetry breaking effect. In Ref. \cite{GLY},
by employing U(3)$_L\times$ U(3)$_R$ chiral theory of mesons, electromagnetic
mass splittings of $\pi$, $a_1$, $K$, $K_1$, and $K^*$ have been calculated
to one loop order and $O(\alpha_{\rm EM})$. In particular,
\begin{eqnarray}
& &(m_{K^+}^2-m_{K^0}^2)_{\rm EM}=0.002473\hspace{0.05in} {\rm GeV}^2=
2 m_K \times 2.5\hspace{0.05in} {\rm MeV},\\
& &(m_{K^{*+}}^2-m_{K^{*0}}^2)_{\rm EM}=-0.003147\hspace{0.05in} {\rm GeV}^2=
-2 m_{K^*}\times 1.76\hspace{0.05in} {\rm MeV}.
\end{eqnarray}
Using the experimental value of mass difference between $K^+$ and $K^0$ \cite{PDG}
together with eqs. (9)-(11), we can get
\begin{equation}
\frac{m_u+m_d}{m_u-m_d}=\frac{m_\pi^2}{(m_{K^+}^2-m_{K^0}^2)_{\rm EXP}-
(m_{K^+}^2-m_{K^0}^2)_{\rm EM}}.
\end{equation}
Then
\begin{eqnarray}
& &m_u+m_d=17.40\pm1.02\hspace{0.05in} {\rm MeV},\\
& &m_u=5.64\pm0.32\hspace{0.05in}{\rm MeV},\;\;\; m_d=11.76\pm0.70\hspace{0.05in} {\rm MeV}.
\end{eqnarray}

From eq. (17), we get the vector meson mass in the chiral limit $m_V$
is $759.5\pm1.0$ MeV. In fact, the value of $2\pi^2 g^2 m$
is about 745 MeV, which is not far from $m_V$. If we assume
$m_V=2\pi^2 g^2 m$, eqs. (17)-(21) will be simplified as follows
\begin{eqnarray*}
& &m_\rho=m_\omega=m_V+\frac{m_u+m_d}{2},\\
& &m_{K^{*\pm}}=m_V+\frac{m_u+m_s}{2},\\
& &m_{K^{*0}}=m_V+\frac{m_d+m_s}{2},\\
& &m_\phi=m_V+m_s,
\end{eqnarray*}
which are different from the corresponding relations given by Urech \cite{UR},
\begin{eqnarray*}
& &m_\rho=m_\omega=m_V+2 \hat{m},\\
& &m_{K^{*\pm}}=m_{K^{*0}}=m_V+\hat{m}+m_s,\\
& &m_\phi=m_V+2 m_s,
\end{eqnarray*}
where $\hat{m}=\frac{m_u+m_d}{2}$.

Taking $m_\phi=1019$ MeV in eq. (21), the value of $m_s$ is about $254$ MeV.
Thus, from eqs. (19) and (20), we can predict
\begin{equation}
m_{K^{*\pm}}=892.13\hspace{0.05in}{\rm MeV},\;\;\;\;\;m_{K^{*0}}=895.27\hspace{0.05in} {\rm MeV}.
\end{equation}
The experimental data from Ref. \cite{PDG} are
\begin{equation}
m_{K^{*\pm}}=891.59\pm 0.24\hspace{0.05in} {\rm MeV},\;\;\;\;\;
m_{K^{*0}}=896.10\pm0.28\hspace{0.05in} {\rm MeV}.
\end{equation}
The small differences between eq. (41) and (42) are due to the electromagnetic
corrections of the $K^*$ mesons. From eq. (41) and eq. (37) [the value of
$(m_{K^{*\pm}}-m_{K^{*0}})_{\rm EM}$], the total mass difference of $K^*$-mesons
is 4.90 MeV, which is in good agreement with the data of
eq. (42) $(m_{K^{*0}}-m_{K^{*\pm}})_{\rm exp}$=4.51$\pm0.52$ MeV.

\section{Summary}

In order to calculate the $\rho^0-\omega$ mixing amplitude,
U(3)$_L\times$U(3)$_R$ chiral theory of mesons is extended to the leading
order in quark mass expansion. The use of path integral method to integrate out
the quark fields naturally leads to the interactions of the $\rho-\omega$
mixing. It has been shown that there is no explicit vector mass-mixing term
in the effective lagrangian of this theory, and the $\rho^0-\omega$ mixing
comes entirely from the quark loop.
The off-shell mixing amplitude is momentum dependent, and vanishes at
$q^2=0$. The on-shell mixing amplitude is obtained from the G-parity forbidden
decay $\omega\rightarrow \pi^+\pi^-$, and the result is in agreement with the
generally accepted value. The current quark masses are extracted from the
on-shell $\rho^0-\omega$ mixing amplitude and the mesons masses. In particular,
the masses of charge and neutral $K^*$-mesons are predicted, and their total mass difference
from both the electromagnetic and isospin symmetry breaking effects
are obtained.

\begin{center}
{\bf ACKNOWLEDGMENTS}
\end{center}
We would like to thank Dr. Xiaojun Wang for reading the manuscript and
pointing out an error in it. One of the authors (D.N. Gao) is particularly
grateful to the referee for providing information on the update analyses of
the $\rho^0-\omega$ mixing in Refs. \cite{OC4,OC5}.
This work is partially supported by NSF of China through C. N. Yang.


\end{document}